\documentclass[sigconf,article]{acmart}   % names shown

\usepackage{booktabs} % For formal tables

\setcopyright{rightsretained}
\usepackage{multirow}

\acmConference[LBRS@RecSys '18]{Late-Breaking Results track part of the Twelfth ACM Conference on Recommender Systems (RecSys'18)}{ October 2-7, 2018}{Vancouver, BC, Canada}

\acmYear{2018}
\copyrightyear{2018}

\acmISBN{}
\acmDOI{} 

\usepackage{flushend}

\begin{document}
%\title{Securing Tag-based recommender systems against profile injection attacks}
\title{Securing Tag-based recommender systems against profile injection attacks: A comparative study}

%The title may also change depending on the experiment result and hopefully praise LSTM.
%\titlenote{Produces the permission block, and copyright information}
%\subtitle{Extended Abstract}
%\subtitlenote{The full version of the author's guide is available as \texttt{acmart.pdf} document}

% COMMENDS BY THE REVIEWERS TO ADDRESS
% 1)
% Summed up, this is a very interesting paper, which is written in a nice way and also the evaluation is sound. Therefore, I vote for accepting it as a poster. For the camera-ready version, I suggest to the authors to proof-read the paper since there are some typos in it.
% 2)
% The details of the DL model are not perfectly clear, especially some design choices, e.g. why do you need two outputs for a two-class problem? % 3)
% Also, it is not clear to me how the sequence model of the LSTM is applied. What is the input sequence and how is there a unique ordering of tags?

\author{Georgios Pitsilis}
%\authornote{Corresponding author}
\affiliation{%
  \institution{Department of Computer Science\\ Norwegian University of Science and Technology (NTNU)}
  \streetaddress{P.O. Box 1212}
  \city{Trondheim}
  \country{Norway}
  \postcode{NO-7491}
}
\email{georgios.pitsilis@ntnu.no}

\author{Heri Ramampiaro}
%\authornote{}
\affiliation{%
  \institution{Department of Computer Science\\ Norwegian University of Science and Technology (NTNU)}
  \streetaddress{P.O. Box 1212}
  \city{Trondheim}
  \country{Norway}
  \postcode{43017-6221}
}
\email{heri@ntnu.no}

\author{Helge Langseth}
%\authornote{}
\affiliation{%
  \institution{Department of Computer Science\\ Norwegian University of Science and Technology (NTNU)}
  \streetaddress{P.O. Box 1212}
  \city{Trondheim}
  \country{Norway}
  \postcode{43017-6221}
}
\email{helge.langseth@ntnu.no}

\renewcommand{\shorttitle}{Securing Tag-based recommender systems against profile injection attacks.}

% COMMENT: I would keep the abstract as clean as possible and avoid using italic ect...
\begin{abstract}
This work addresses challenges related to attacks on 
social
%collaborative 
tagging systems, which often comes in a form of 
malicious annotations or profile injection attacks. In particular, we study various countermeasures against two types of threats for 
such
%social tagging
systems, the Overload and the Piggyback attacks.
The studied countermeasures 
%we studied 
include baseline classifiers such as, Naive Bayes filter and Support Vector Machine, as well as a deep learning-based approach. Our evaluation performed over synthetic spam data, generated from del.icio.us, shows %
that in most cases, the deep learning-based approach provides the best protection against threats.
%%%% COMMENT (Heri): I changed to all cases because the results showed that DL is better than the others in all our cases...

%\emph{Social Bookmarking} and \emph{Tagging}, also called {\em Collaborative Tagging}, has emerged a new era in user collaboration. Collaborative Tagging allows users to annotate resources of their liking and keep organized, using one or more personalized tags. With the application of suitable algorithms, these collaborative environments can be useful in the provision of resource recommendations. %Nevertheless, such systems are still vulnerable to attacks which come in the form of malicious annotations. 
%A challenge is, however, that such systems are often vulnerable to attacks in a form of malicious annotations or profile injection attacks.
%In this paper, we study various countermeasures against two types of such attacks for social tagging systems, the \emph{Overload attack} and the \emph{Piggyback attack}. The tested countermeasure schemes include baseline classifiers such as, \emph{Naive Bayes filter} and \emph{Support Vector Machine}, as well as a \emph{Deep Learning} approach. Our evaluation performed over synthetic spam data generated from \emph{del.icio.us} dataset, shows %
%that in most cases, Deep Learning can outperform the classical solutions, providing a high 
%that the above countermeasures can still provide a substantial 
%level of protection against threats.

%.\footnote{This is an abstract footnote}
\end{abstract}



%\ccsdesc[500]{Information systems~Collaborative and social computing systems and tools}
%\ccsdesc[500]{Security and privacy~Web application security}


%\renewcommand\footnotetextcopyrightpermission[1]{}
%\settopmatter{printacmref=false}
%\setcopyright{none}

\maketitle

\section{Introduction}
Recommender Systems (RS) are information filtering mechanisms, aiming at predicting the preference of users to particular resources.
%\emph{Collaborative tagging},
%a web-based service that is representative of the new Web 2.0 technology,
%allows users to store, share and annotate various kinds of web resources
%(e.g news, photos)
%, such as news, blogs, and photos
%into social data repositories.
%Also known as 
\emph{Social Tagging} systems, 
%or \emph{Social Bookmarking} systems, 
%such tools 
facilitate resource recommendations using the users' assigned annotations to resources, (aka \emph{folksonomies}), as input to
%various 
prediction algorithms.
%In a social annotation system, the assignment of tags to %resources by users is referred to as
%expressed by structures called 
%\emph{folksonomies}
%, which are tuples of four components: $F=(U,T,R,A)$, where $U$,$T$ and $R$ are finite sets of users, tags and resources, respectively, and $A$ denotes a ternary relationship between them, e.g., $A\subseteq U\times T\times R$. A tag assignment $a=(u,t,r) \in A$ is an annotation of resource $r$ by user $u$ with tag $t$.
%
%The novelty of annotating using textual tags 
The novelty of using tags for annotation
%
%The novelty of tags
%tags (a.k.a textual annotations) as source of %information
%
%With tags (a.k.a textual annotations) being a novel source of information, their concept
has attracted the interest of scientists
in RS
%with a literature rapidly %expanding (\cite{Fernquist:2013,Heymann2008,Haustein2012}).
%expanding 
\cite{Fernquist:2013,Heymann2008},
%maybe need o be replaced by newer citations.
%\input{samplebody-conf}
as well as of attackers.
%, aiming to benefit by promoting or disapproving a product.
%However, like any RS,
%the open structure of RS have
%The open structure and the adaptive characteristics of RS have, however, 
%they are vulnerable to various types of attacks,
%aiming
%In general, the aim of an attack to a RS is 
%to promote or disapprove a product for the benefit of the attacker.
%
%A common way for an attacker to achieve this is to inject malicious profiles that correspond to fictitious identities into the system, biasing it towards recommending inferior products to users. 
%Likewise to ordinary rating-based Collaborative-filtering systems, \emph{collaborative tagging} RS have also attracted the interest of attackers.
%Like other types of recommender systems,
%\emph{collaborative tagging} has attracted the interest of attackers, aiming at distorting the system's behavior. Such an attack can be compared to attacking an ordinary rating-based Collaborative-filtering system in tag-based recommender systems.
%In tag-based recommender systems, there is an intention by the attackers that is similar with attacking an ordinary rating-based Collaborative-filtering system.
%The existence of these
% **********************************
In general, an attack against a tagging system consists of 
%a set of
%one or many 
coordinated malicious profiles that correspond to fictitious identities, injected into the system,
for
biasing the recommendation algorithm towards suggesting inferior products to users.
%Such threats highlight the need for 
%effective countermeasures for
%safeguarding tag-based RS.
%with each one being associated with a fictitious user identity. 
%Their annotation history 
%The annotation history of such an identity
%is designed so that to bias the recommendation algorithm.

% **********************************

The issue of security in tag-based RS
%in not new, but it 
has so far been mainly approached 
%by solutions associated with spam detection.
%
using 
%as being similar to spamming.
%
%Our focus 
%instead
%in this work 
%is on 
%the so called \emph{Identification-based} solutions, which aim to 
%detecting and isolating any potentially threatening 
%entities,
%users or resources.
%such as a user or a resource.
%As per \cite{Heymann:2007}, anti-spam approaches in social networks are divided into 3 main categories: a) \emph{Prevention mechanisms}, such as CAPTCHAs \cite{VonAhn2003}, b) \emph{Rank-based} approaches which demote spam in search queries, and c) \emph{Identification-based} which aim to detect and isolate any potentially threatening entities, such as a user or a resource. 
%The focus of our work is on the latter type only.
%
%To filter out non-legitimate users, 
%Various
anti-spamming techniques
%have been 
%developed by numerous researchers in the field, 
%employing either a form of 
%based on 
such as, \emph{Bayesian} type filtering~\cite{Yazdani2012}, or other tag classification methods~\cite{Koutrika:2007}. 
In the above works, particular characteristics of the tags
used in annotations are exhibited, 
%such as the presence of suspicious ones into user profiles. In the above works%
%the assumption made is 
with the assumption that
%any 
tags used by legitimate users would coincide with each other.
%those by 
%other legitimate users.
Nevertheless, such 
%spam tag 
filtering becomes ineffective if attackers are aware of the 
%correlations being used in an 
attack filtering policy
%and thus use non-suspicious tags
~\cite{Koutrika:2007,Yazdani2012}. 
%as they can
%they might try to
%disguise their fake profiles by using non-suspicious tags~\cite{Koutrika:2007,Yazdani2012}. 
%In such a case, 
Other feature-based countermeasures,
%works that detect spam by exploiting users' features, 
employ either the neighbors' honesty within the group \cite{Zhai2016}, or mix features of tags and users together along with other, derived from the social connectivity \cite{Poorgholami}.
%
%Previous works that exploit users' features to build a spam detection mechanism, employ either the neighbors' honesty within the group \cite{Zhai2016}, or mix features of tags and users together along with other, derived from the social connectivity \cite{Poorgholami}.
%
\iffalse
\citet{Zhai2016} have developed a spam detection mechanism based on the user-behavior, that employs criteria such as the neighbors' honesty within the group, of which the user is member.
A similar filtering approach by \citet{Poorgholami} mixes features of tags and users together in the spam classification, that include the spamicity for tags from Baysian classification along with some user features derived from the social connectivity.
\fi
%
Neural network-based approaches, including \emph{deep learning}~(DL),
%are approaches that
are known to provide 
%a 
good 
%level of 
protection 
against
%for the general task of spam detection 
spamming~\cite{gauri2017}.
%in emails
%and in social posts
%
%%% COMMENT/QUESTION (Heri): Reading from this sentence: DL provide good level of protection against both email spamming and social posts. Is this what you really want to tell? What is "social posts" in this setting, btw?
%
%Deep learning (DL) has been proved quite effective on data classification, taking advantage of the large amounts of data and the growth in the available computational power.
%
However, 
%despite the plethora of solutions existing today for spam classification, there is still a need for alternative and modern approaches, such as DL. 
%
%we strongly believe that 
%Furthermore, the existing solutions, both those based on supervised or unsupervised learning, are not tailored for protecting tag-based RS, generic 
%
%Despite that, the existing solutions, either based on supervised or unsupervised learning models, attempt to address the problem of spamming in folksonomies with the provision of generic solutions. As such not enough consideration is taken into .... personalized recommendations.
%
%As far as the employment of Deep Learning techniques in the task of securing a tag-based Recommender System from the injection of malicious profiles, to the best of our knowledge, ...no work has been done towards investigating the effectiveness of deep learning for the purpose of detecting malicious annotations.
%
%Furthermore, there is no study today to provide quantifiable results concerning the effectiveness of the detection mechanisms expressed in terms of impact in the accuracy on the personalized recommendations received.
%
%To
%to the best of our knowledge, 
DL has neither been adequately investigated so far
%as a good solution to prevent
for %preventing
safeguarding
%attack prevention in 
tag-based RS, 
%and no 
nor
%have 
quantifiable results 
%have been provided 
exist to show
%demonstrate 
%how 
the effectiveness of the countermeasures 
%is reflected to
on the recommendations.
%output.

% **************************************

%There are generally many commonalities between attacking a tag-based RS and the phenomenon of spamming. 
%The main focus of this work is on special forms of attacks with interesting characteristics associated with a tag-based RS, as found in the literature~\cite{Ramezani2009}.

% **************************************

%With this in mind, 
The research question
%addressed
%that 
we address
%in this work 
is:
\emph{How effective are the various classifier schemes 
%in the battle 
against profile injection attacks,
%the injection of bogus profiles, 
%whose aim is 
that aim
to influence the personalized recommendations %received by users 
in a tag-based 
%recommender system?
RS?}
%
% maybe we revise it depending on the experiment output.
To answer this question, our main goals are:
%can be summarized as:
%into the following:
\emph{i)} to study various classification schemes,
%including \emph{deep learning}, 
over known types of shilling attacks, and \emph{ii)} to evaluate the
effectiveness of classification
%success of classification, as well as 
%its effectiveness
into the recommendation process.
%the actual effectiveness of filtering-out any bogus folksonomies.
%\begin{itemize}
%\item To study various classification schemes, including \emph{deep learning}, over known %types of shilling attacks.
%\item To evaluate the success of classification, as well as the actual effectiveness of %filtering-out any bogus folksonomies.%, expressed in performance metrics for tag-based recommender systems.
%\end{itemize}

Our contribution in this paper is two fold: 
%consisting of:
%and provides:
%First,
\emph{i)} a synthetic set 
%dataset 
of 
%consisting of
malicious
data
%annotations
to
serve
%be used for
testing purposes, 
%Second, 
%\emph{ii)} A comparative study of the effects of known attacks in a typical personalized tag-based RS, demonstrating the effectiveness of potential countermeasures against such attacks. Those include a \emph{deep learning} (DL) model adapted to 
%the requirements of 
%the issue we come to address.
and \emph{ii)} a comparative study of the effects of known attacks and the effectiveness of potential countermeasures against them, in a typical 
%personalized 
tag-based RS. Those include a
properly adapted
DL model.
%adapted to 
%the requirements of 
%the issue we come to address.
% ****************************************************

\iffalse
The rest of the paper is organized as follows:
In Section~\ref{preliminaries}, we describe in more detail 
the 
%problem of 
attacks in tag-based RS.
%with focus on the types of attacks we are addressing. 
In Section~\ref{relWork}, we discuss the existing work in the field. 
In Section~\ref{Approach}, we present our proposed solutions;
%to the issue; 
while in Section~\ref{Evaluation}, we describe the dataset we used, 
out evaluations,
%the evaluation performed, 
and discuss 
%the 
our 
results.
%received. 
Finally, in Section~\ref{Conclusion}, we summarize our contributions and outline the future work.
\fi

%\section{Problem Statement}
%\section{Preliminaries}
%\label{preliminaries}
%\subsection{Problem Statement}
%\label{Statement}

%\subsection{Attacks Description}
\label{Attacks}

\vspace*{-0.06in}
\section{Securing the folksonomies}
\label{Approach}
In this work, we focus on two forms of intrusions known as \emph{Overload} and \emph{Piggyback} attacks 
%with interesting characteristics associated with 
%for
%tag-based RS
~\cite{Ramezani2009}. The goal of the former is to overload a tag context with a \emph{bogus} resource to achieve correlation between the tag and that resource. To accomplish this, an attacker associates the \emph{bogus resource} with a number of popular tags. For the latter, the objective is the \emph{bogus resource} to ride the success of another highly \emph{popular}
one.
%resource, we call 
%by associating them in a way to appear similar.
To achieve that, an attacker would annotate the \emph{bogus} resource choosing any popular tags already associated with the \emph{popular} one,
so that they appear similar.
%In this section 
%We 
%briefly describe the three supervised learning classification algorithms 
%that
%we included in our comparative study.
Our comparative study includes the following algorithms:
%We describe the classification algorithms of our comparative study.

\noindent\textbf{Naive Bayes filtering:} 
%\paragraph{Naive Bayes filtering}
%Naive Bayes 
%Its effectiveness 
%It is known to be very effective
%for binary classification of emails 
%for detecting
%in order to detect 
%spam, is quite known.
%For the needs of our attack detection mechanism
%In our case, we applied the  well-known formula derived from the Bayes theorem,
%To detect attacks, we applied Naive Bayes filtering 
%for classifying folksonomies, based on the existence of tags in them.
%and we
%More specifically, we
%used the
%~\cite{Sahami98abayesian}.
%\subsection
%\paragraph
%
It is a quite known classifier for detecting spam emails based on the Bayes theorem, here applied
%To detect attacks, we applied Naive Bayes filtering 
for classifying folksonomies based on the existence of tags in them.
%and we
%More specifically, we
%used the
%~\cite{Sahami98abayesian}.
%\subsection
%\paragraph

\noindent\textbf{Support Vector Machine (SVM):} 
%We 
%It was included 
%SVM 
%in our study 
%mainly because of its suitability 
%as being suitable
%for binary classification.
%tasks.
%The reason for the inclusion of SVM into our comparison is mainly the suitability of this method for binary classification of data.
%We chose the \emph{linear kernel} function for SVM for 
%classifying
%separating 
%the input folksonomies into 
%two classes, 
%\emph{legitimate} and \emph{malicious}.
%To apply this, 
%on our dataset, 
%Before the classification the input folksonomies were vectorized and %then
%they were 
%transformed into TF-IDF values,
%The latter 
%transformation 
%was necessary 
%to scale down the impact of the most frequent tags
%found
%in the dataset.
%
%Prior to that, were vectorized and transformed into TF-IDF values
%to scale down the impact of the most frequent tags.
%found
%in the dataset.
The input folksonomies were first vectorized and transformed into TF-IDF values
to scale down the impact of the most frequent tags.
Then, they were classified into 
%two classes, 
\emph{legitimate} and \emph{malicious},
using the \emph{linear kernel} function.
% for SVM.

\vspace*{-0.09in}
\setlength{\abovecaptionskip}{10pt}
\setlength{\belowcaptionskip}{-05pt}
%\begin{figure}[th]  %[!htbp]  %[!ht]
\begin{figure}[!hbp]  %[!ht]
\centering
%\hspace*{-0.25in}
%\includegraphics[width=.49\textwidth]{lstm_model_CR}
\includegraphics[width=.49\textwidth,height=.21\textwidth]{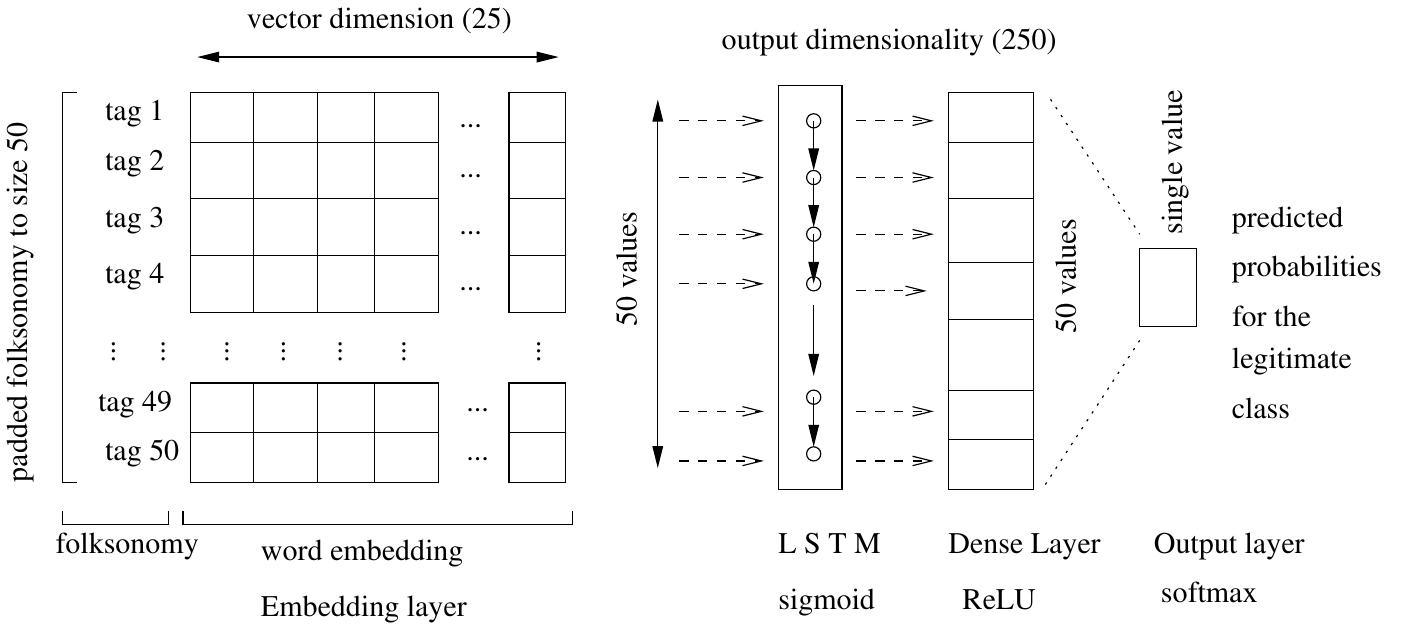}
%\vspace*{-0.25in}
%\caption{High level view of our deep learning model}
\caption{The deep learning model architecture used}
\label{fig:model}
\end{figure}
%\section{Evaluation}
%\vspace*{-0.29in}
\vspace*{-0.05in}
\noindent\textbf{Deep Learning (DL):} We used a hybrid classification model which employs a Long-Short-Term-Memory (LSTM)-based Recurrent Neural Network and
%(RNN).
%algorithm. 
%This hybrid model 
%has the strong benefit of working 
works
both with and without sequential data. It is composed of four layers (see Fig.\ref{fig:model}): \emph{a)} an 
%\emph{input} layer (a.k.a
\emph{Embedding} layer
%)
with size relevant to the vocabulary of tags used in the folksonomies (set to 25),
\emph{b)} a \emph{hidden} layer, fully connected to the input and the subsequent layer, with dimensionality of its outer space set to 250, 
based on preliminary results,
\emph{c)} a \emph{dense} layer, used for improving the learning and stabilize the output, of size equal to the input folksonomy, and
\emph{d)}
a single neuron \emph{output} layer provides the classification output in the form of probabilities for the \emph{Legitimate} class.

\section{Experiments}
\label{Evaluation}

%In order t
To demonstrate the effects of the attacks on the recommendations, %output,
we employed \emph{Vector Space},
a legacy algorithm for personalized recommendations in tag-based RS,
which we adapted to our particular case.
%model in our experimentation, a legacy recommendation algorithm for tag-based 
%RS~\cite{Salton1975}.
%for personalized recommendations
%that provides personalized recommendations of resources to users
%~\cite{GEMMELL2012}.
%In this one, the distance between a user and resource vector is computed out of
%by taking into account 
%the frequencies of tags associated with them.
%As the name implies, vector space similarity is employed for computing the distance between a user's and an resource's vector, taking into account the frequencies of tags associated with them.
%For our 
%particular
%case, we adapted 
%the above 
%this
%model to the needs of
%folksonomy-based 
%\emph{Social Tagging} systems.
%Each user is represented by a multinomial feature vector $u=[w_{t_1},w_{t_2},...,w_{t_n}]$ of dimension $n$, equal to the number of different tags in the vocabulary of the words used, with $w_t$ denote as the weight of the particular tag $t$ on that user. Vector weights $w_t_i$ may be expressed through many ways, with the frequency of tags being the most common.
First, we express every tag in the corpus as \emph{word2vec} vector
%~\cite{Mikolov2013}, 
%extracted from
by \emph{Google's} pre-trained set. % by 
Next, every 
user-posted
folksonomy 
%posted by a user 
is represented by a single vector, 
%computed 
by averaging the \emph{word2vec} values of all tags in that folksonomy.
%For representing each user by a single vector, first we express every tag of every folksonomy by that user as a \emph{word2vec} \cite{Mikolov2013} vector, selected from a pre-trained set by \emph{Google}.
%Next, a folksonomy vector is computed by averaging the word2vec values of all tags in the folksonomy. 
%In that way,
As such, every user, or resource acquires its own vector representation, 
%derived 
by combining together all folksonomy vectors associated with 
them.
%that user / resource.
%the vector representation of that user's profile is expressed as: $u=[u_{t_1},u_{t_2},...,u_{t_n}]$, over a set of $n$ tags : $t_1,t_2,..,t_n$, by combining together all folksonomy vectors by this user.
%In a similar way, resources are also modeled as vectors, over the same set of $n$ tags as: $r=[r_{t_1},r_{t_2},...,r_{t_n}]$.
%Finally, 
%the \emph{Cosine Similarity}
%the similarity 
%of a user's vector and a resource
%between the vectors of a user and a resource
%can be 
%derived
%worked out
%by 
%computing 
%the \emph{Cosine Similarity} value between their vectors.
%In such a way, recommendations of resources to users can take the form of personalized lists, containing the \emph{top-k} resources in terms of their cosine similarity value with that user.
%In such a way,
%As such, 
%is used for building the 
%the personalized \emph{top-k} lists of recommended resources for each user.
%are built based on the their \emph{Cosine Similarity} with that user.
Finally, the personalized \emph{top-k} recommendations for a user is build using the \emph{Cosine Similarity} of his vector and the resourses' vectors.

%The cosine value for each pair of vectors can be derived from the Euclidean dot product formula, given in eqn.\ref{eqn:cosine}, with
% $\vec{t_1} \cdot \vec{t_2}$ denote as the inner product of the two vectors, and $||\vec{t_n}||$ is the
%length of vector $t_n$.

%( ... common knowledge - to be removed ... )

%\begin{equation}
%cos(\vec{t_1},\vec{t_2})=\frac{\vec{t_1} \cdot \vec{t_2}}{
%||\vec{t_1}|| \ ||\vec{t_2}||}
%\label{eqn:cosine}
%\end{equation}

\vspace*{-0.06in}
\subsection{Dataset and Evaluation Metrics}
\vspace*{-0.03in}
%We chose to 
%do our evaluation
%evaluate our hypothesis
%on 
%using annotation data from 
%\emph{del.icio.us} dataset.
%\footnote{Dataset acquired from the work in \cite{basile_topical_2015}.}, 
%a public 
%of folksonomies.
%To reduce processing time, 
For every run
in our evaluation
%we restricted the data 
%to average users who have annotated 20 or less items, corresponding in total to a subset of 25 thousand users and 62k folksonomies.
%To maintain balance 
we selected randomly subsets 
%of folksonomies 
%each corresponding to
of
3k users from \emph{del.icio.us} dataset, 
corresponding to
%The resulting subsets are composed of 
%around 
73k folksonomies
that form corpuses of 42k different tags.
Due to lack of pre-labeled bogus data,
we built synthetic bogus folksonomies of size and tag content determined
%
%we used synthetic bogus folksonomies, the size of which as well as the tags selected, were determined 
%
according to 
guidelines
%that we 
found in the literature~\cite{Ramezani2009}. %concerning the \emph{Overload} and \emph{Piggyback} attacks
%\footnote{To retain anonymity, the dataset with the bogus profiles will be made publicly available after the completion of the review process.}.
As such, to simulate the \emph{Overload} attack, the tags of the fake folksonomies were chosen out of the 75 most popular ones used in the legitimate folksonomies, while the max size of the fake
ones
%folksonomies
was limited to 50 tags.
The actual 
size of
%number of tags in
a fake folksonomy was chosen so that, legitimate 
and fake ones will follow the same distribution.
The popular tags were selected from those used for annotating the most popular resources.
%folksonomies in the original dataset. so that the fake and legitimate ones to be identical.
%To determine the actual number of tags to include in a fake folksonomy, we studied the distribution of the number of tags in the legitimate folksonomies in the original dataset, and we build bogus ones with size that follows the same distribution.
%In addition, for the \emph{Piggyback} attack, 
%due to dataset sparsity 
%of \emph{del.icio.us},
%the tags used for building 
%the fake folksonomies were build choosing tags
%chosen 
%from 
%those used in 
%the top-100 most annotated resources.
%in the dataset. 
%This value was chosen as being suitable, given the sparsity of \emph{del.icio.us} dataset.
%\begin{table}%[!htbp]
%   \centering
  % \caption{Statistics of the dataset (\emph{del.icio.us}) used in our evaluation.}
%  \caption{Statistics of our \emph{del.icio.us} dataset.}
%   \label{tab:stat_dat}
%   \scalebox{0.99}{ 
%   \begin{tabular}{lc}
%   \toprule
   %\hline
%   \textbf{Statistics of dataset} & \textbf{\emph{del.icio.us}} \\
   %\hline
%   \midrule
%   \emph{\# folksonomies} & 73k \\
   %\hline
%   \emph{\# users}        & 3k \\
   %\hline
%   \emph{\# unique tags (corpus)}  & 42k \\
   %\hline
%   \bottomrule
%   \end{tabular}}
%\end{table}
%\subsection{Evaluation Metrics}
The impact of the attacks is demonstrated via
%To demonstrate the impact of attacks, 
%we chose 
%to adopt 
a set of approved metrics
%from the literature 
\cite{Ramezani2009}
we adopted, which are:
%For the \emph{Overload} attack we used:
%the following:
1) the \emph{F-Score} for the spam classification,
2) the \emph{Avg. rank} 
of % appearance of
%in which 
the bogus resource
%appears
in the users' \emph{top-k} lists, and
%of recommended resources.
3) the \emph{population} affected by the attack, 
%i.e., the number of 
(users been recommended a bogus resource).
%
%\begin{itemize}
%    \item \emph{F-Score} for the spam classification process.
%    \item \emph{Average rank} in which the bogus resource appears in the users' %\emph{top-k} lists of recommended resources.
%    \item The \emph{population} affected by the attack, i.e., the number of users for %which the bogus resource has been recommended.
%\end{itemize}
%Especially
For the \emph{Piggyback} attack, the last metric refers to
users
%of users 
for whom the bogus resource has been ranked higher than the popular one.
%in 
%the
%their \emph{top-k} recommendations.
%they received.

%In addition, for the Piggyback attack we used the following:

%\begin{itemize}
%    \item Affected Population, that is the users for whom the attack item has been ranked higher than the popular item in their received recommendations.
%    \item Average Rank of Popular Item in the users' recommended top list.
%    \item Average Rank Improvement, that is the number of places the bogus item has been ranked above the popular item in the users' top lists. That means, the smaller the value, the more resistant the countermeasure is.
%\end{itemize}

\vspace*{-0.06in}
\subsection{Experimental Setup}
\vspace*{-0.03in}
For training the classifiers 
%over the bogus class
%of bogus data, a 
%realistic and yet suitable for demonstrating the effects of the attacks,
we appended 30\% fake synthetic folksonomies onto the set
%total number
of legitimate ones.
%fix ratio of fake synthetic folksonomies was appended to the dataset, corresponding to 30\% of the total number of the legitimate ones.
%This value was chosen as being as much as realistic for sufficiently training the classifiers and yet making possible to appropriately demonstrate the effects of the attacks.
For the testing, we chose variable \emph{attack size},
%We used the \emph{attack size} as a variable
ranging from 0.1\% to 10\%,
%for
%in 
%the testing, 
%of each classifier. 
%and it
%This variable
which
refers to the ratio of the 
%injected 
fake folksonomies
%into the system,
over 
the
%the total number of
legitimate ones.
%folksonomies, and it ranged from 0.1\% to 10\%.
%Every 
%experimental 
%setup was run 5 times, and 
%then 
%the results were averaged.

To demonstrate the effectiveness of each %classification 
algorithm,
for each setup,
fake folksonomies
and 
legitimate ones
were mixed together and supplied into the vector space model to
compile
the \emph{top-k} 
recommendations ($k$ was set to 15)
for each user,
both before and after applying the countermeasures.
For all three filtering algorithms we tested, we performed 10-fold cross validation over the sample data. 
%in which the 9 folds were used for training and the remaining one for testing the algorithm on unseen data.
%We implemented 
The DL model was implemented in the Keras
%\footnote{https://github.com/fchollet/keras}
toolkit, 
applied
%used the 
{ADAM} optimization
%algorithm
\citep{KingmaB14}
and selected
\emph{categorical cross-entropy} as the learning objective.
%$The corpus size was set to 42k for our dataset, which is the number of different tags used in the sample folksonomies.
Also,
%Furthermore, 
we modeled
%chose to model
the input folksonomies (Fig.~\ref{fig:model}) in the form of vectors using word-based frequency vectorization.
%This means that
%In other words, 
%the tags in the corpus were indexed based on their frequency of appearance in the corpus and the index value of each 
%tag
%word
%was used as vector element to describe the folksonomy.
Finally, the DL model was
trained
%allowed to run 
for an 
optimal
number of epochs.
%%% COMMENT (Heri): How many runs exactly?

%until reaching an optimally trained state. 
%That state is reached when the validation accuracy is maximized, while at the same time the error remains within $\pm 1\%$ of the lower ever figure within that fold in the cross validation.

\vspace*{-0.06in}
\subsection{Results and Discussion}
\begin{figure}[!ht]
\centering
\hspace*{-0.25in}
%\vspace*{-1.1in}
%\includegraphics[width=0.56\textwidth]{overload_delicious.pdf}
%\includegraphics[width=0.56\textwidth]{all_delicious.pdf}
%\includegraphics[width=**cm,height=**cm,angle=0]{filename}
%\includegraphics[scale=0.33]{all_delicious.pdf}
\includegraphics[height=160pt,width=290pt]{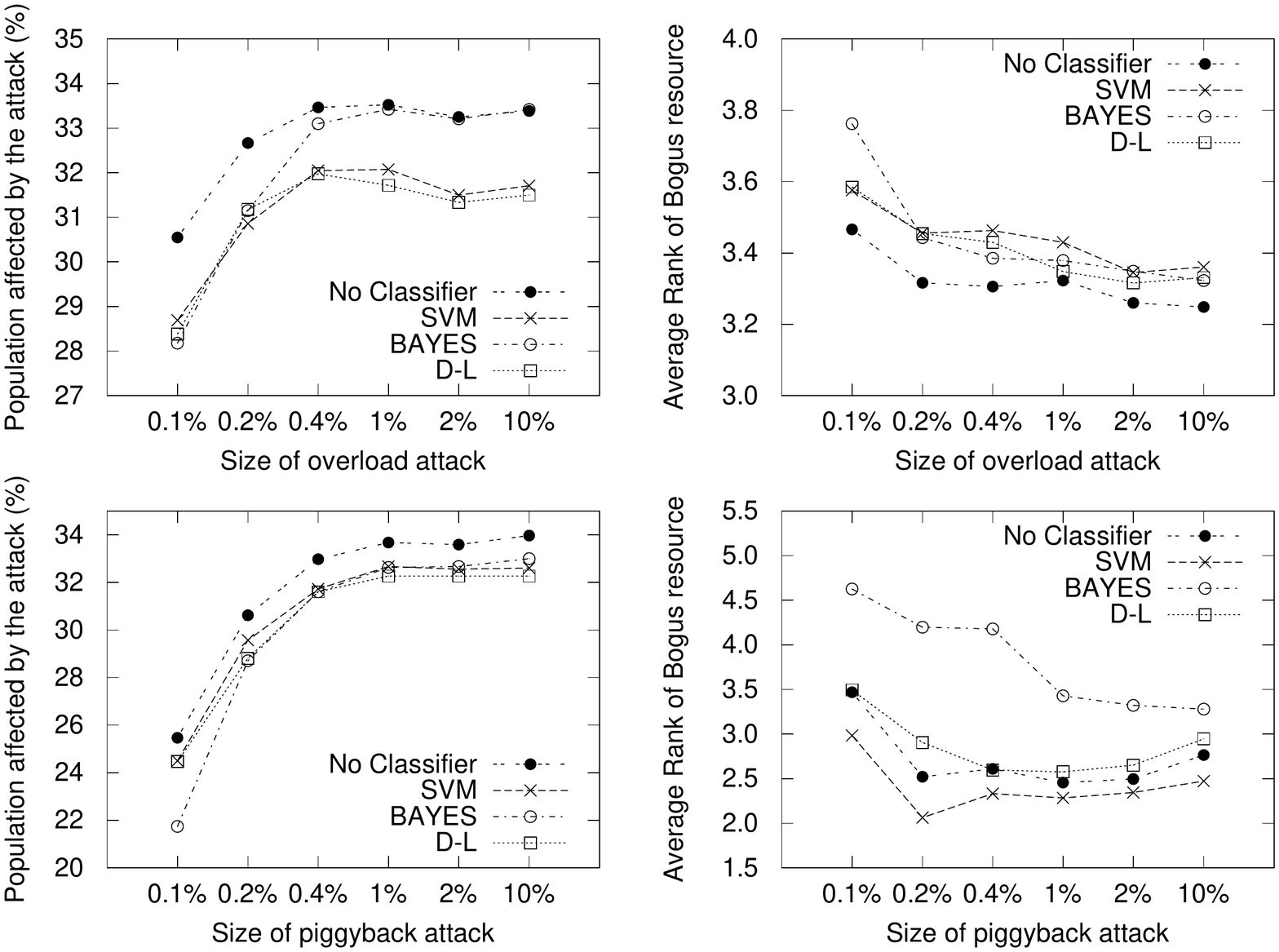}
\vspace*{-0.13in}
\caption{
%For both
%Overload \& Piggyback 
%attacks, is shown t
The affected population (small values indicate strong resistance), and the rank of bogus resource (large values indicate strong resistance)}
\label{fig:graph_ol}
\end{figure}

\vspace*{-0.03in}
The results are the average values of five runs.
%We present the most interesting results of our experimentation.
%As can be seen in Table~\ref{tab:Results_ov_del},
%showing the classifiers' F-Score performance, 
%the 
DL
%approach
outperforms the other approaches (see Table~\ref{tab:Results_ov_del}) in classifying
%both 
the 
legitimate and 
%the 
bogus folksonomies, for both attacks.
As far as 
%the performance in 
the recommendation service
(see
%as can bee seen in
%shown in
%the population affected by the attack is shown in the top parts of
Fig.~\ref{fig:graph_ol}),
%\& 
%Figure~
%\ref{fig:graph_pg}
%while the rank of bogus resource, shown in the bottom parts of these figures refers to any users 
%who have been 
%affected by the attack.
%as can be seen
very
%Very 
interestingly, even attacks of small scale are enough to render a significant population of users vulnerable.
In fact, the DL approach, in comparison to the other alternatives, provides in general, good resistance
%in preventing the intrusion 
to intrusions
of bogus resources into the users \emph{top-k} lists, 
for
%This is something 
%That seems to be the case of 
both
%types of 
attacks.
Also,
%very interestingly
%is the fact that 
in terms of 
the Avg. Rank of the Bogus resource (See Fig.~\ref{fig:graph_ol}, right), DL scales better for large sizes of attacks 
vs
%as compared to 
the Bayes classifier, 
but
%which seems to 
performs best for small attacks only. 
For the same metric in the piggyback attack, DL despite being the second best performing, it 
also does better
%the result improves with the increase of the
for large
attacks,
%attacks size,
as opposed to Bayes classifier. %for which this is diminishing.
%\begin{table*}[ht]
%\begin{table*}[ht]
%\setlength{\abovecaptionskip}{0pt}
%\setlength{\belowcaptionskip}{0pt} 
\begin{table}[!t]
   %\vspace{-2mm}
   \centering
   \caption{Classification accuracy for both attacks}
   \label{tab:Results_ov_del}
   \vspace*{-0.12in}
   \scalebox{0.9}{ 
   \begin{tabular}{|c|c|c|c|c|c|c|}
   \hline 
    & \multicolumn{3}{|c|} { Overload } & \multicolumn{3}{|c|} { Piggyback } \\
   \hline 
   \multicolumn{1}{|c|} { F-score }  & \textbf{SVM}   & \textbf{BAYES}  & \textbf{DL} & \textbf{SVM}   & \textbf{BAYES}  & \textbf{DL} \\
   \hline
   %\cline{1-7}
    overall &  0.9501 &	0.8339	& \bf{0.9570}   & 0.9680	 & 0.9009 &	\bf{0.9728} \\
   \hline
   legitimate  & 0.9665 & 0.9082 &	\bf{0.9709} & 0.97888 & 0.93878 & \bf{0.9818} \\
   \hline
   bogus  & 0.8958 & 0.5863 & \bf{0.9104} & 0.9319	 & 0.7748 & \bf{0.9426} \\
   \hline
   \end{tabular}}
\vspace{-0.20in}
\end{table}

%
%\begin{figure}[!ht]
%\centering
%\hspace*{-0.3in}
%\vspace*{-0.2in}
%\includegraphics[width=0.56\textwidth]{F-Scores-delicious.pdf}
%\vspace*{-0.25in}
%\caption{F-score achieved for the \emph{del.icio.us} data}
%\label{fig:graph_F_delicious}
%\end{figure}
%
%\begin{figure}[!ht]
%\centering
%\vspace*{-0.4in}
%\hspace*{-0.3in}
%\includegraphics[width=0.56\textwidth]{F-scores-citeUlike.pdf}
\vspace*{-0.28in}
%\caption{F-score achieved for the \emph{citeUlike} data}
%\label{fig:graph_F_citeUlike}
%\end{figure}
%\setlength{\abovecaptionskip}{0pt}
%\setlength{\belowcaptionskip}{0pt}
\setlength{\abovecaptionskip}{-10pt}
\setlength{\belowcaptionskip}{-0pt} 
\vspace*{-0.20in}
\section{Conclusions and Future Work}
\label{Conclusion}
In this paper, we investigated the impact of spam filtering
%on resource recommendation
in tag-based RS.
%To simulate 
We simulated
two known attacks, 
by generating 
%we generated synthetic
fake 
data
%folksonomies
from original, 
taken 
from \emph{del.icio.us}. 
%dataset.
%
%.... 
%Very surprisingly, filtering models which achieved the highest F-score in the attack filtering process, have finally failed to do so in .... , were finally resulted to have negative impact on the actual prediction accuracy in the resource recommendation process. 
%.....
Our experiments showed that our deep learning model outperforms all the legacy classifiers in terms of F-score and, 
in most cases, it can safeguard the user recommendations.
%having also a scalability advantage.
% in the overload attack only.
%For larger attacks it can potentially outperform, in all aspects, the legacy classifiers.
%The benefits of D-L as opposed to other supervised learning approaches for classification....
%For future work, we plan to 
Our future work includes
%further improve the filtering performance by investigating
%investigate other deep learning architectures and 
%experiment
experimentation
with feature extraction from folksonomy data to feed the neural network,
as well as  generalizing our results by exploring more datasets.

%In addition, we intend to further generalize the results by exploring more 
%resource annotation
%datasets.
%, as well as performing extensive comparative studies.% against more resource recommendation algorithms.
\bibliographystyle{ACM-Reference-Format}

\bibliography{tagging}

\end{document}